\documentclass[talk1,a4paper,fleqn]{w-art1}
\usepackage{times,w-thm1,amsmath,amsfonts}

\theoremstyle{plain}

\theoremstyle{definition}

\newcommand{\ket}[1]{{|#1 \rangle}}

\newcommand{\bea}{\begin{eqnarray}}
\newcommand{\eea}{\end{eqnarray}}
\renewcommand{\bar}{\overline}

\newcommand{\Tr}{\hbox{Tr}}

\renewcommand{\tilde}{\widetilde}

\usepackage[]{graphicx}

\begin{document}

\pagespan{1}{}

\hspace{11cm} {\tt NSF-KITP-09-60}\\

\title[Emergent geometry in ${\cal N}=6$ Chern-Simons-matter theory]
{Emergent geometry in ${\cal N}=6$ Chern-Simons-matter theory}

\author[Diego Trancanelli]{Diego Trancanelli  
\footnote{ E-mail:~\tt{dtrancan@physics.ucsb.edu}}}
\address
{Department of Physics and Kavli Institute for Theoretical Physics \\ University of California at Santa Barbara,
CA 93106, USA }

\begin{abstract}

We investigate a strong coupling expansion of ${\cal N}=6$
superconformal Chern-Simons theory obtained from the semiclassical
analysis of low energy, effective degrees of freedom given by the
eigenvalues of a certain matrix model. We show how the orbifolded
sphere $S^7/\mathbb{Z}_{k}$ of the dual geometry emerges dynamically
from the distribution of the eigenvalues. As a test of this approach
we compute the energy of off-diagonal excitations, finding perfect
agreement with the dispersion relation of giant magnons.

\medskip

\noindent {\it Talk given at the ``BIRS Workshop on Gauge Fields,
Cosmology, and Mathematical String Theory'', Banff (Canada), Feb. 1
- 6, 2009.}

\end{abstract}
\maketitle


One of the most intriguing aspects of the AdS/CFT correspondence is
the fact that the strong coupling dynamics of the gauge theory
encodes information about classical gravitational physics and the
geometry can be seen as {\it emerging} from the saddle points of the
gauge theory path integral. A remarkable example of this phenomenon
is the emergence of the sphere of the dual $AdS_5 \times S^5$
geometry from the matrix quantum mechanics governing the scalars of
${\cal N}=4$ super Yang-Mills compactified on a $S^3$
\cite{B}.\footnote{For a review of subsequent developments and a
list of relevant references see \cite{Brev}.}

Following \cite{Berenstein:2008dc}, we investigate the notion of
emergent geometry in the  context of the three-dimensional ${\cal
N}=6$ superconformal Chern-Simons theory that was proposed in
\cite{ABJM} as the world-volume action for coincident M2-branes.
This theory consists of a $U(N)\times U(N)$ Chern-Simons gauge
field, with levels $k$ and $-k$, coupled to four chiral superfields
transforming in bifundamental representations of the gauge group. We
call them $(\phi^A)^i_{\; i'}$, with $A$ an $SU(4)$ R-symmetry index
ranging from 1 to 4 and $i,i'=1,\ldots,N$.\footnote{In the following
we will denote with the same symbol $\phi^A$ also the lowest
components of these superfields.} We will also use later the
notation $\phi^A=(A_a,\bar B_{\dot a})$, with $a,\dot a=1,2$, so
that $A_a$ transforms in the $(N,\bar N)$ while $B_{\dot a}$ in the
$(\bar N,N)$. The theory enjoys a large $N$ limit and one can also
define a 't Hooft coupling constant, given by $\lambda\equiv N/k$.
For $k \gg N$ the theory is weakly coupled, while for $N\gg 1$ and
$k\ll N$ it has a gravity dual which is either M-theory on
$AdS_4\times S^7/\mathbb{Z}_k$ (if $k\ll N^{1/5}$) or type IIA
strings on $AdS_4\times \mathbb{C}P^3$ (if $N^{1/5}\ll k\ll N$).
Taking $k$ large shrinks in fact the radius of the M-theory circle
fibered over the $ \mathbb{C}P^3$ base, thus reducing the dimensions
of the space from eleven to ten.

For our purposes, we are interested in the gauge theory living on
$\mathbb{R}\times S^2$. We will show how one can consistently
truncate the fields to just the constant modes of the scalars on the
$S^2$ and reduce everything to a (gauged) matrix quantum mechanics.
The eigenvalues of this model represent effective degrees of freedom
that dominate the low energy physics. The semiclassical quantization
of these modes will induce an effective interaction, forcing the
eigenvalues to localize on a 7-sphere. This 7-sphere has to be
regarded as the sphere in the dual geometry and it will also have a
preferred direction which we will identify with the M-theory circle.
As a consistency check for these identifications, we will compute
the dispersion relation of giant magnons from this approach, finding
perfect agreement with the results in the integrability literature.


The starting point for our analysis is to consider the states in
chiral ring, {\it i.e.} the states with equal energy and R-charge. 
Using the state/operator correspondence, one can show that
these states correspond to field configurations that are spherically
symmetric on the $S^2$.

This condition of spherical symmetry implies that the gauge field
needs to be covariantly constant on the sphere and reduces to
constant magnetic fluxes along the Cartan directions of the gauge
group. We break then  $U(N)\times U(N)\to U(1)^N \times U(1)^N$,
with fluxes ${\cal F}^i$ and $\tilde {\cal F}^{i'}$ for the two
factors. These fluxes are topologically quantized already at the
classical level \cite{Atiyah:1982fa}. The scalars are charged under
the gauge group and experience a magnetic monopole background of
strength ${\cal F}^i-\tilde {\cal F}^{i'}$. When expanding in
spherical harmonics on the sphere one has thus to consider monopole
spherical harmonics \cite{Wu:1977qk}, which have angular momentum
given by $\ell+|m_i-\tilde m_{i'}|/2$. Spherical symmetry
corresponds to $\ell=0$, namely $s$-waves on the sphere, and
$m_i=\tilde m_{i'}$, thus inducing an identification between the
fluxes ${\cal F}^i$ and $\tilde {\cal F}^{i'}$. In particular the
gauge group $U(1)^N \times U(1)^N$ is further broken down to a
diagonal subgroup $U(1)^N$.

Moreover, one can show that the violation of the BPS bound for these
configurations is given by a sum of squares that are essentially the
F- and D-term constraints. Imposing that these constraints vanish
confines the fields to the moduli space region of normal and
commuting matrices. Finally, one has to satisfy the equation of
motion for the (temporal component of the) gauge field, which acts
as a Lagrange multiplier enforcing Gauss' law. This equation is
given by \bea k {\cal F}^i = (\phi^A\dot{\bar\phi}_A-\dot\phi^A
\bar\phi_A)^i_i\equiv Q_i\,, \eea with a similar expression applying
to the tilded quantities. Given the quantization of the fluxes, we
see then that also these charges are quantized at the classical
level and are only aligned along the Cartan directions. Putting all
these ingredients together we can recognize that the moduli space of
the M2-branes is given by $Sym^N(C^4/\mathbb{Z}_k)$, as computed
also in \cite{DMPV,ABJM}.


In summary, imposing that the configurations we are looking at
correspond to states in the chiral ring leaves us with four normal
and commuting $N \times N$ matrices which represent the $s$-waves of
the scalars on the $S^2$. We call these modes $\phi^A$ as well. They
have the following action \bea {\cal S}= \int dt\, \Tr\left(
\dot{\bar\phi}_A \dot\phi^A -\frac{1}{4}\bar\phi_A \phi^A + {\cal
V}_6\right)\label{action}\,, \eea where the mass term originates
from the conformal coupling of the scalars to the curvature of the
sphere and ${\cal V}_6$ is a sextic potential given by \bea {\cal
V}_6 &=& \frac{1}{12 k^2}\left[\phi^A \bar\phi_A\phi^B
\bar\phi_B\phi^C \bar\phi_C +\bar\phi _A \phi^A \bar\phi_B \phi^B
\bar\phi_C \phi^C \right.\cr && \hskip 2cm\left. +4\,\phi^A
\bar\phi_B\phi^C \bar\phi_A \phi^B \bar\phi_C -6\,\phi^A
\bar\phi_B\phi^B \bar\phi_A\phi^C \bar\phi_C\right]\,.\label{V6}
\eea

Since the matrices are normal and commuting we can diagonalize them
simultaneously with a gauge rotation \bea \phi^A = U^{-1} D^A V\,,
\eea where $U$ and $V$ are unitary, time dependent transformations
acting, respectively, on the unprimed and primed indices of the
bifundamental fields $\phi^A$. The matrix $D^A$ is diagonal, with
eigenvalues $x^A_i\equiv \vec x_i$.

We need now to compute the Jacobian factor due to the
change of basis from generic matrices to diagonal ones, {\it i.e.}
the volume of the gauge orbit. One way to do it is to start from the
kinetic term for the off-diagonal degrees of freedom of $\phi^A$ and
read off the determinant from the induced metric on the space of
these modes. Expanding $\phi^A$ (here $\theta$ and $\tilde \theta$
are the gauge parameters of the $U$ and $V$ gauge transformations
and we suppress the $SU(4)_R$ indices for a moment) \bea (\dot
\phi)^i_{\; i'}= \dot x^i_{i'}-i \dot\theta^i_{\; j}x^j_{i'}+i
x^i_{j'}\dot{\tilde\theta}^{j'}_{\; i'} \,, \qquad x^i_{i'}\equiv
x_i \delta^i_{i'}\,, \eea one finds (setting for simplicity $N=2$)
\begin{equation}
\Tr\, ( {\dot { \bar \phi}}_A \dot \phi^A )=
\begin{pmatrix}
\dot{ \theta}^2_1 & \dot {\tilde \theta}^2_1
\end{pmatrix}
\begin{pmatrix}
|x _1|^2 +|x _2|^2
&-2x _1  x^* _2 \\
-2 x^* _1x _2 &|x _1|^2 +|x _2|^2
\end{pmatrix}
\begin{pmatrix}
\dot \theta^1_2\\
\dot{\tilde \theta}^1_2
\end{pmatrix}\, .
\end{equation}
The determinant of this matrix (generalized to arbitrary $N$) is the
volume of the gauge orbit \bea \mu^2 = \prod_{i<j} \left[\left(|\vec
x_i|^2+|\vec x_j|^2\right)^2-4|\vec x_i \cdot \vec
x^*_j|^2\right]\,. \label{mu2}\eea Notice how $\mu^2$ is gauge
invariant and symmetric in the exchange of two eigenvalues. It is
moreover invariant under the independent rephasing of single
eigenvalues \bea \vec x_i \to e^{i \alpha_i} \vec
x_i\,.\label{resid-gauge} \eea We will return soon to discussing the
meaning of this residual gauge invariance.

Taking into account the determinant factor, the Hamiltonian for the
eigenvalues $\vec x_i$ following from (\ref{action}) is given by
\begin{equation}
H = \sum_i -\frac 1{2 \mu^2} \vec\nabla_i \cdot \mu^2 \vec\nabla_i
+\frac 18 |\vec  x _i|^2\label{eq:heff}\,.
\end{equation}
The insertion of the quantum measure $\mu^2$ in the kinetic term for
the eigenvalues induces a semiclassical quantization for these
degrees of freedom.

It is easy to see that \bea \psi_0=\exp\left(-\frac{|\vec x
_i|^2}{4}\right) \label{eq:grst} \eea is an eigenfunction of $H$.
This in fact turns out to be the wave function of the ground state. It is
convenient now to absorb the volume of the gauge orbit into the wave
function by defining $\hat \psi_0~\equiv~\mu  \psi_0$. This wave
function is symmetric in the exchange of eigenvalues, which are then
to be interpreted as bosons. With this choice, the measure of
integration for $\hat\psi_0$  is just the standard
Euclidean measure $\prod_i d\vec x_i$. The probability density for
the eigenvalues in the ground state of the system is given by \bea
|\hat \psi_0|^2 =
 \exp\left( -\frac 12 \sum_i |\vec x _i|^2+ \frac 12\sum_{i,j}\log
 \left[( |\vec x _i |^2+|\vec x _j|^2)^2- 4 |\vec x _i \cdot\vec{ x }^{\, *}_j|^2\right]\right)\,.
 \label{prob}
\eea If we go to the thermodynamic limit, we have a continuous
distribution of eigenvalues and we can turn the sums into integrals
by introducing an eigenvalue density, $\sum_i \to \int d\vec x \,
\rho(\vec x)$. One finds that this density has to have a singular
support compatible with the symmetries of the theory and reads
\bea \rho(x)=\frac{3 N}{\pi^4 r_0^7}\delta(|\vec x|-r_0)\, . \eea
The value of the radius $r_0$ can be found by plugging $\rho(x)$
into the expression for the probability (\ref{prob}) and by
maximizing the exponent. One readily finds that \bea |\hat \psi_0|^2
= \exp \left( -\frac N 2 r_0^2 +2 N^2\log r_0+ N^2 c\right)\, , \eea
where $c$ does not depend on $r_0$. From this it follows that
\begin{equation}
r_0= \sqrt{2N}\, .
\end{equation}
We see then that there is a competition between the attractive force
due to the harmonic oscillator potential given by the mass term and
the repulsive force coming from the Jacobian determinant. The
equilibrium configuration of the bosonic eigenvalue gas corresponds
to a uniform distribution on a 7-sphere of radius $r_0$, which we
claim has to be identified with the sphere of the dual geometry.

To make this identification more precise we need to introduce the
off-diagonal modes that have been integrated out so far. The basic
idea is to introduce these modes as perturbations around the
classical configuration of commuting matrices, ignoring the
back-reaction. The important quantity to compute is their mass,
which is obtained by expanding the sextic potential ${\cal V}_6$ in
(\ref{V6}) to quadratic order. After a rather tedious computation
one finds that the mass of the off-diagonal mode connecting the
$i$-th and $j$-th eigenvalues is given by
 \bea m^2_{ij}
=\frac{1}{4}+\frac{1}{4k^2}\left[ \left(|\vec x_i|^2+|\vec
x_j|^2\right)^2-4|\vec x_i \cdot \vec x^*_j|^2 \right]\, .
\label{mass-od} \eea We see then that this mass scales like the
distance squared between the two eigenvalues, a fact that has to be
contrasted with what happens in ${\cal N}=4$ super Yang-Mills, where
the scaling is linear in the distance \cite{B}. Notice moreover that
we recover the same dependence as in the Jacobian determinant
(\ref{mu2}). This fact is not too surprising, after all introducing
the off-diagonal modes corresponds to higgsing the gauge group down
to $U(1)^N$. Recalling that the eigenvalues live on a sphere of
radius ${\cal O}(\sqrt{N})$, one sees also that the mass
(\ref{mass-od}) scales like $\lambda=N/k$, so that at strong
coupling it was consistent to integrate out these degrees of freedom
and the low energy physics was indeed dominated by the diagonal
modes.

The fact that the mass of off-diagonal modes goes like a distance
(squared) between the eigenvalues gives a heuristic argument for the
emergence of a notion of locality. The eigenvalues are in fact
positions on the sphere (which is dense in the $N\to \infty$ limit),
and points far away from each other are disconnected, because the
corresponding off-diagonal modes are heavy and can be integrated
out.

So far we have seen the emergence of a 7-sphere. We can actually be
more precise and realize that there is a special direction in this
sphere. The Jacobian (\ref{mu2}) and the mass
(\ref{mass-od}) are in fact invariant under the symmetry
(\ref{resid-gauge}). This means that the repulsion between the
eigenvalues is not sensitive to their position along the circle
corresponding to this symmetry, and, equivalently, the off-diagonal
modes cannot discriminate between points along this
circle.\footnote{For this reason, locality along the M-theory circle
is much more difficult to establish and has to arise in a
non-perturbative fashion \cite{Hosomichi:2008ip}.} This suggests
that this direction has to be identified with the M-theory circle
that is the natural Hopf fiber of the $S^7$ over $ \mathbb{C}P^3$.
Remembering moreover that the typical size of an eigenvalue scales
like $\sqrt{N}$, we observe that the mass of the off-diagonal modes
scales like \bea m\sim \sqrt{N}\, \frac{\sqrt{N}}{k}\,, \eea {\it
i.e.} it scales like an area rather than like a length, so that
these modes have to be two-dimensional surfaces rather than strings.
This is a way to understand that the theory proposed in \cite{ABJM}
is really a theory of M2-branes and not of D2-branes. The factor
$1/k$ shows moreover that these off-diagonal modes, or {\it membrane
bits}, have to wrap the M-theory circle. The effect of the orbifold
is in fact to shrink the M-theory circle by a factor of $1/k$, so
that for $k\to \infty$ this circle disappears, the theory reduces to
type IIA string on $AdS_4\times  \mathbb{C}P^3$, and the membrane
bits reduce to string bits.


To provide further evidence that the sphere emerging from the
distribution of eigenvalues should be really identified with the
sphere in the geometry, we compute now the energy of (massive)
string excitations given by the off-diagonal modes and compare the
result with the dispersion relation of the giant magnon solution on
$AdS_4 \times  \mathbb{C}P^3$ found in \cite{Nishioka:2008gz,Gaiotto:2008cg,Grignani:2008is}.
The approach we have described so far will in fact allow  in a very simple
way for a complete resummation of certain Feynman diagrams to all
orders in the world-sheet momentum $p$ of the giant magnon. This
resummation will give from first principles the exact square root
functional form conjectured in \cite{Nishioka:2008gz,Gaiotto:2008cg,Grignani:2008is}
assuming integrability. The agreement we find 
shows that the emergent sphere has the correct size in string
units.

The main idea \cite{Berenstein:2005jq} of the following computation
is to turn on the off-diagonal modes $\delta\phi^A$
 neglecting their back-reaction on the geometry. This is allowed as long as these modes are heavy enough.  They
can then be approximated as  free harmonic oscillators with
Hamiltonian \bea H =\sum_{i\neq j'} (\Pi_A)^i_{\;
j'}(\Pi^{A*})^{j'}_{\; i} + m^2_{ij'} (\delta\phi^A)^i_{\;
j'}(\delta\phi^*_A)^{j'}_{\; i} \, , \label{Hho} \eea where $\Pi_A$
is the conjugate momentum of $\delta\phi^A$ and the frequency is
given by (\ref{mass-od}).

To compute the energy  of the BMN operators we recall that the
ground state of the (alternating) spin chain of ${\cal N}=6$
Chern-Simons is $\Tr(\ldots A_1B_1A_1B_1 A_1 B_1\ldots)$
\cite{Minahan:2008hf}, and $A_1$ has eigenvalues $x^1_i$ while $B_1$
has eigenvalues $x^*_{3 i}$. A 2-impurity state is given by \bea
\ket{\psi} = \sum_{l =0}^J e^{2\pi iq l/J} \sum_{i,i'}(x^1_i x^*_{3
i})^l (\Phi^\dagger)^i_{\; i'}( x^{1}_{i'}x^*_{3  i'})^{J-l}
(\Phi'^\dagger)^{i'}_{\; i}  \ket{\hat\psi_0} \, , \label{psi} \eea
where we have called $\Phi$ and $\Phi'$ the impurities in the spin
chain and $2\pi q/J=p$. The impurities are given by replacing in the
combination $(A_1B_1)$ either $A_1$ with $A_2$ or $\bar B_2$ or by
replacing $B_1$ with $B_2$ or $\bar A_2$
\cite{Minahan:2008hf,Gaiotto:2008cg}. We  interpret these
off-diagonal impurities as raising operators for the Hamiltonian
(\ref{Hho}). The energy of the state $\ket{\psi}$ is then given by
\bea \left< E\right>_\psi=\frac{\int (\prod dx_i)|\hat \psi_0|^2
\sum_{i,i'} \big|\sum_l e^{2\pi iql/J}(x^1_i x^{*}_{3 i})^l
(x^{1}_{i'} x^*_{3 i'})^{J-l}\big|^2    \, m_{ii'}}{\int (\prod
dx_i)|\hat \psi_0|^2 \sum_{i,i'}\big|\sum_l e^{2\pi iql/J}(x^1_i
x^*_{3 i})^l (x^1_{i'} x^*_{3 i'})^{J-l}\big|^2  }\, ,
\label{energy0} \eea where we have used that the lowering and
raising operators obey $ [(\Phi)^i_{\;i'},(\Phi^\dagger)^{j'}_{\;
j}] = \delta^i_j \delta^{j'}_{i'}$, and similarly for the primed
operators.

The integrals in (\ref{energy0}) can be evaluated with a saddle
point approximation in two steps. First of all, the integrals are
going to be dominated by the configurations which maximize $|\hat
\psi_0|^2$, corresponding to eigenvalues that are distributed on the
sphere of radius $r_0=\sqrt{2N}$. We
take $N$ to be large, so that the distribution of eigenvalues is
continuous.  In this regime, the sum over two particles can be
replaced by integrals over the sphere. Next we should
maximize $|x^1 x_3^*|$. This  clearly requires $|x_2|=|x_4|=0$.
We can use the gauge freedom (\ref{resid-gauge}) to choose the
following convenient parametrization for a pair of eigenvalues that
satisfy these constraints \bea \vec x=r_0(e^{i\varphi}\cos\vartheta
,\,0,\,\sin\vartheta,\,0)\, ,\qquad \vec
x'=r_0(e^{i\varphi'}\cos\vartheta' ,\,0,\,\sin\vartheta',\,0)\, .
\eea Maximizing the expressions $(x^1 x^*_3)^l$ and $(x'^1
x'^*_3)^{J-l}$  requires then  that  $\vartheta=\vartheta'=\pi/4$.
We can at this point rewrite the mass of the off-diagonal modes in
(\ref{mass-od}) in a more explicit way as \bea
m^2=\frac{1}{4}+\frac{1}{k^2}\left(r_0^4-\left|\vec x \cdot \vec
x'^*\right|^2\right)=\frac{1}{4}+4\lambda^2\sin^2\left(\frac{\varphi-\varphi'}{2}\right)\,
, \eea where, again,  $\lambda=N/k$.

Putting all together, we get
 \bea \left< E\right>_\psi=\frac{\int
d\varphi\,d\varphi' \, \big|\sum_l \exp\left(2\pi iql/J+i l \varphi
+i (J-l) \varphi'\right)\big|^2  \sqrt{\frac 14+4\lambda^2
\sin^2(\varphi-\varphi')/2}}{\int d\varphi \,d\varphi' \,
\big|\sum_l \exp\left(2\pi iql/J+i l \varphi +i (J-l) \varphi'
\right)\big|^2  }\, . \label{energy} \eea Expanding the modulus
squared of the sum over phases one obtains \bea \sum_{l,l'=0}^J \exp
\left[ i(l-l')\left(\frac{2\pi q}{J}+\varphi-\varphi'\right)\right]
&=&\frac{\sin^2(J+1)\left(\frac{2\pi
q}{J}+\varphi-\varphi'\right)/2}{\sin^2\left(\frac{2\pi
q}{J}+\varphi-\varphi'\right)/2}\,. \eea In the limit of large $J$,
this is sharply peaked for $ \varphi-\varphi'=2\pi q/J$, and acts as
a delta function. The energy reduces then to \bea \left< E
\right>_\psi=\sqrt{\frac{1}{4}+4\lambda^2 \sin^2\frac{p}{2}}\, ,
\label{disp-rel-fin} \eea in perfect agreement with the result
obtained in \cite{Nishioka:2008gz,Gaiotto:2008cg,Grignani:2008is} using
integrability. Notice that we get the weak coupling value of the
interpolating function $h(\lambda)$
\cite{Nishioka:2008gz,Gaiotto:2008cg,Grignani:2008is} under the square root. This is
because we have used the value of the $r_0$ computed from the
weak-coupling expression for the volume of the gauge orbit. At
strong coupling the radius of the sphere can undergo a (finite)
renormalization \cite{BHH} and the functional dependence on
$\lambda$ inside the square root will change.  The emergent geometry
picture guarantees nevertheless that the square root behavior will
hold to {\it all orders} in perturbation theory in the
renormalization of the radius of the distribution and, therefore,
also for $\lambda\gg 1$. This renormalization does not take place in
${\cal N}=4$ super Yang-Mills, where the dispersion relation for the
giant magnon is exact. This can be understood using S-duality
arguments \cite{BTprep}.

The relation (\ref{disp-rel-fin}) is actually exact also with
respect to another quantum number. We can in fact consider higher
harmonics of the scalars on the $S^2$, corresponding to bound states
of giant magnons \cite{Berenstein:2007zf}, and replace the $1/4$ in
the expression (\ref{disp-rel-fin}) with $(\ell +1/2)^2$, where
$\ell$ can be interpreted as the momentum along the $AdS_4$
directions. Giant magnon solutions of the string sigma model
\cite{Ahn:2008hj} show agreement with this result.


\section*{Acknowledgements} I am grateful to David Berenstein for
collaboration on these topics and for many illuminating discussions.
I would also like to thank the organizers of the  ``BIRS Workshop on
Gauge Fields, Cosmology, and Mathematical String Theory'' as well as
all the participants for creating a very stimulating meeting. This
work was supported in part by the U.S. Department of Energy under
grant DE-FG02-91ER40618 and by the National Science Foundation grants PHY05-51166 and PHY05-51164.


\end{document}